%% file: main.tex
\documentclass{llncs}
\usepackage[utf8]{inputenc}
\usepackage[backend=biber,maxbibnames=9, maxcitenames=6, maxalphanames=4, minalphanames=3, giveninits=true,style=alphabetic]{biblatex} %, style=alphabetic
\usepackage{amsfonts}
\usepackage{amssymb}
\usepackage{graphicx}
\usepackage{verbatim}

\usepackage{algorithm}
\usepackage{graphicx}
\usepackage[leqno]{amsmath}
\usepackage[noend]{algpseudocode}
\usepackage{booktabs}
\usepackage{appendix}
\usepackage{threeparttable}
\usepackage{tikz}
\usepackage{enumitem}
\usepackage[framemethod=tikz]{mdframed}

\usepackage{cleveref}
\usepackage{fullpage}

\mdfsetup{skipabove=0pt,skipbelow=0pt}

	\newcommand{\balgo}{\begin{algorithm}}	\newcommand{\ealgo}[2]{\end{mdframed}

\caption{#1}\label{#2} %toujours le label aprÃ¨s le caption ! sinon Ã§a renvoie au paragraphe courant 
\end{figure}
\end{algo}}

\newenvironment{hugenum}{\begin{description}[topsep=0ex,itemsep=0ex,partopsep=1ex,parsep=1ex, font=\sffamily\bfseries] }{\end{description}}
\newenvironment{grandenum}{\begin{enumerate}[topsep=0ex,itemsep=0ex,partopsep=1ex,parsep=1ex, label=\arabic*)] }{\end{enumerate}}
\newenvironment{petitenum}{\begin{enumerate}[topsep=0ex,itemsep=-1ex,partopsep=1ex,parsep=1ex, label=(\roman*)] }{\end{enumerate}}
\newcommand{\genum}{\begin{grandenum}} \newcommand{\fenum}{\end{grandenum}} \newcommand{\senum}{\begin{petitenum}} \newcommand{\esenum}{\end{petitenum}}

 \usepackage{fancyhdr}
\makeatletter %for enabling vertical separator: [cc|cccc]
\renewcommand*\env@matrix[1][*\c@MaxMatrixCols c]{%
  \renewcommand*{\arraystretch}{.7} % for reducing row space
  \hskip -\arraycolsep
  \let\@ifnextchar\new@ifnextchar
  \array{#1}}
\makeatother

\newtheorem{defi}{Definition}
\newtheorem{main}{Main Theorem} \newtheorem{prop}{Proposition}

\newcommand{\A}{\mathcal{A}} \newcommand{\quer}{M} %number of queries

\newcommand{\F}{\mathbf{F}} 
 
\newcommand{\co}{\mathcal{O}}

 \newcommand{\sake}{\mathrm{SAKE}}
\newcommand{\ti}{\widetilde}
\renewcommand{\vec}{\overrightarrow}

\newcommand{\bc}{\mathsf{BC}}

\newcommand{\key}{\ensuremath{K}} 
\newcommand{\keys}{{\ensuremath{|K|}}}
\newcommand{{\iv}}{\ensuremath{{IV}}}

\newcommand{\s}{S} 
\newcommand{\nbs}{\ensuremath{n}} %number of shares
\newcommand{\nbst}{m} %number of shares required for reconstruction, the threshold

\newcommand{\x}{X} %input vector 
\newcommand{\y}{Y} %output vector
\newcommand{\xlen}{x} %input vector length
\newcommand{\ylen}{y} %output vector length

\newcommand{\blen}{\ensuremath{{|B|}}} 

\newcommand{\plain}{\ensuremath{P}} %PLAIN
\newcommand{\pbl}{P} %a plaintext block
\newcommand{\plainl}{\ensuremath{|\plain|}} %size of the plaintext in bits
\newcommand{\plen}{\ensuremath{{\ensuremath{c-1}}}} %length of the plaintext 

\newcommand{\ciph}{\ensuremath{C}} %CIPH
\newcommand{\cbl}{C} %a ciphertext block
\newcommand{\clen}{\ensuremath{c}} %number of blocks in the ciphertext
\newcommand{\ciphl}{|\ciph|} % ciphertext size in bits
\newcommand{\perm}{\sigma}
\newcommand{\enc}{Enc} %encryption scheme
\newcommand{\trans}{AON} %{\mathcal{S}} %transformation (e.g. SSAKE)
\newcommand{\ra}{\rightarrow} \newcommand{\lra}{\longrightarrow}
\newcommand{\prng}{\ensuremath{\{0,1\}^{\blen}  } } %{PRNG()}

\newcommand{\caon}{{SSAKE}} \newcommand{\kecchare}{ROSSake}
\renewcommand{\comment}[1]{} %UNCOMMENT TO HIDE commentaire

\providecommand{\keywords}[1]{  \small	  \textbf{\textit{Keywords---}} #1 }

\begin{document}
\title{Revisiting Shared Data Protection Against Key Exposure}
\author{Katarzyna Kapusta
\inst{1}\thanks{Work done while at Télécom Paris}
\and Gerard Memmi\inst{2} \and Matthieu Rambaud \inst{2}
}
\institute{Thales, cortAIx \email{katarzyna.kapusta@thalesgroup.com} \and LTCI, Télécom Paris, Institut Polytechnique de Paris \email{matthieu.rambaud, gerard.memmi@telecom-paris.fr}}

\maketitle

{\centering
	\makebox[0.98\linewidth]{Version 2 - May 2026\footnotemark }
\setcounter{footnote}{3}
\footnotetext{Updated analysis of Bastion, clarifications in the proof and update of related work.}
}

\begin{abstract}
This paper puts a new light on computational secret sharing with a view towards distributed storage environments. It starts with revisiting the security model for encrypted data protection against key exposure. The goal of this revisiting is to take advantage of the characteristics of distributed storage in order to design faster key leakage resisting schemes, with the same security properties as the existing ones in this context of distributed storage.

We then introduce two novel schemes that match our ---all storage places or nothing--- security level of secret sharing under key exposure. The first one is based on standard block cipher encryption. The second one fits both in the random oracle model (e.g. Keccak) or in the idealized blockcipher model with twice larger key than the security parameter (e.g. an idealized AES256 would achieve 128 bits security). The first one reduces by half the amount of the processing required to be done in addition to data encryption with regard to the fastest state-of-the-art solution, whereas the second one completely eradicates additional processing. We confirm the complexity results by presenting a performance evaluation.

A non-negligible part of our contribution lies in the provided security analysis. In addition to giving security proofs for our schemes, we revisit the ones of a previous scheme and point out a gap in the proof when the key is exposed.
We propose a hardness assumption which looks reasonable and which enables to bridge the gap.
\end{abstract}

% \begin{CCSXML} %code generated by the tool at https://dl.acm.org/ccs.cfm
% <concept_id>10002978.10002979</concept_id>
% <concept_desc>Security and privacy~Cryptography</concept_desc>
% <concept_id>10002978.10003006.10003013</concept_id>
% <concept_desc>Security and privacy~Distributed systems security</concept_desc>

% \ccsdesc[500]{Security and privacy~Cryptography}
% \ccsdesc[500]{Security and privacy~Distributed systems security}


\pagenumbering{gobble}

\section{Introduction}

\input{introduction.tex}

\input{data_structures_and_notations.tex}

\input{relevant_work.tex}

\input{security_model.tex}

\input{ssake_description.tex}

\input{security_analysis.tex}

\input{ROSake.tex}

\input{comparison.tex}

\input{conclusions.tex}

% % \bibliographystyle{ACM-Reference-Format} %may switch to abbrv to gain space
% % \bibliography{sample-bibliography}
\printbibliography

\input{appendix.tex}

\end{document}

%% file: introduction.tex
Ensuring confidentiality of data at rest is mostly achieved by its encryption using a symmetric cipher. However, even the strongest algorithm will protect data only as long as the key remains secret to attackers. Secure key management is the obvious countermeasure to that problem but its good implementation is not so straightforward and may very well be very costly. The problem of reinforcing data confidentiality against weak or leaked keys exists in the literature since almost three decades \cite{bib:rivest}. Nowadays, the emergence of new powerful attackers puts this problem into a new light  \cite{bib:bastion,QKLQM19,trustcom}.

As data are often outsourced and managed externally, key exposure origins may be manifold. First, it may be the result of bad key generation, for instance because of the use of hard-coded keys or backdoors in the key generation software~\cite{duhk,bib:bastion}. Second, the key may be revealed because of its poor management (the risk of bad key management increases in a situation where multiple users share same data and the same key). Third, with the time passing, the key length may become not sufficient anymore for always more powerful adversaries acquiring enough computational capabilities. Last but not least, an encryption key may simply be lost, stolen, or obtained as the result of bribery or coercion.

A classic way of reinforcing encrypted data confidentiality and protecting it against key exposure consists in encrypting the data in such a way that an adversary obtains no information unless she has the totality of the ciphertext, plus a secret encryption key: this property is denoted as \textit{all-or-nothing} (AON). Key exposure resilient secret sharing schemes can thus be derived from schemes with this property. However, we observe that former \textit{all-or-nothing} works (as the classical Rivest's or Desai's schemes \cite{bib:rivest}) are \emph{not efficient} because they require two rounds of encryption: the first one to achieve the all-or-nothing property, and the other one to encrypt. Also, a more recent work \cite{bib:bastion} gets rid of this twice-encryption requirement, at the cost of a linear post-processing of the encrypted file.
But it appears that its security proof is based on the assumption that an adversary could not distinguish between a piece of a ciphertext (encrypted in counter mode)---of which she chose the plaintext--- and pure randomness. 
We give more details in \Cref{sub:cexctr}, where we show that this assumption does not hold in the adversary model considered by both \cite{bib:bastion} and us, which is that the adversary is given the encryption key.
We then introduce, in \Cref{sub:fixbastion}, an assumption that bridges the gap in their proof and which looks reasonable.

\textit{How can we \emph{efficiently} protect encrypted data against key exposure, since we believe that the risk of key exposure is impossible to completely eradicate ?}

By \emph{efficiently} we mean with no additional encryption layer. We discuss the general technical challenges raised by this problem throughout our paper: in §\ref{sub:krawbreak}, at the end of \ref{remark-on-r}, in §\ref{sub:nail} and in §\ref{sec:revisiting-assumptions}. The good news is that distributed storage provides us with an additional security ingredient - data fragmentation and dispersal - enabling us to address the key exposure threat. On the other hand, it seemed to us that all the previous schemes address an \emph{over constraining} security model, where the adversary selects all the ciphertext blocks that she wants to see (up to a certain quantity). Guided by a practical secure storage purpose, we think that the relevant parameter is instead \emph{the number of} shares containers or storage sites that an adversary is able to access to. Therefore, another question: \textit{In a threat model where an adversary fully accesses a certain number of storage sites, can we achieve the \emph{same security level} as the existing AON schemes with \emph{more efficiency}?}

As will become clear, we are facing a tradeoff. Either we could follow the approach of Bastion~\cite{bib:bastion} and use mainstream encryption, then fix the issue of non-pseudorandomness by an additional post-processing. Or we could strive to design encryption schemes that \emph{directly} have a pseudorandom behavior under key exposure:

\textit{Can we find \emph{efficient} encryption schemes where any large fraction of the ciphertext looks pseudo-random even in a situation of key exposure? And for mainstream encryption, how far can we reduce the necessary post-processing overhead to achieve this goal?}

We address the second question by revisiting the security model for encrypted data protection against key exposure in a distributed environment such as cloud composed of several storage sites or multi-cloud with independent storage providers.
Instead of taking the previous AON threat model (described in  \cite{bib:bastion}) in which an adversary is able to compromise a given number of \emph{any} ciphertext \textit{blocks}, we present a new model taking as parameter the number of compromised \textit{servers or sites}. Such a change is motivated by the fact that an adversary able to compromise one ciphertext \textit{block} in a given storage site has necessarily been able to acquire enough access rights to also be able to compromise other data stored in the same site by the same user, containing the said compromised ciphertext \textit{blocks}.
In particular, \emph{we match the same security level} as the recently introduced CAKE~\cite{bib:bastion} model, in our context where the adversary accesses full shares.

This re-adjustment of the security model allows us to propose new secret sharing schemes with the all-or-nothing property that are faster than the state-of-the-art propositions. First, we introduce a scheme based on classical symmetric encryption with a block cipher in the counter mode. It reduces by half the complexity of the post-processing when compared with the fastest relevant work, which is Bastion's scheme~\cite{bib:bastion}. We show then how to totally get rid of the post-processing by switching to a scheme based on less mainstream encryption with random oracles. We give two efficient instantiations of it: from ideal blockciphers with twice larger keys (e.g. AES256 would provide a security parameter of 128), and from ideal sponge functions (e.g. Keccak). The price to pay for our new schemes improvements is a slight storage overhead, of $\nbs-1$ times the keysize (of e.g. 128 bits, for an overall 128 bits level of security), where $\nbs$ is the number of storage sites.

Apart from complexity and performance improvements, the main contribution of this work is a deep security analysis of the computational secret sharing schemes in a situation of key exposure. We revisit the previously proposed security proof of \cite{bib:bastion} and single out a sufficient hypothesis under which its security is guaranteed. At the same time, we evidence that the classic proof structure for indistinguishability of plaintext encryptions, which proceeds by reduction to a distinguisher between an implemented keyed function and an idealized one, breaks down in presence of key leakage. This thus motivates the completely different approach in our proofs, which crucially rely on the fact that an idealized blockcipher with leaked key, still behaves as an ideal permutation. To match this ideal behavior in practice, it is important that the leaked key should have been chosen sufficiently at random in the first place. The proof of our alternative scheme in the random oracle model is actually much simpler, and provides a more efficient scheme.

In conclusion, our work comes as both two scalable and fast secret sharing schemes, and a complement to key management recommendations provided, for instance, by the NIST~\cite{B16}.

\textbf{Detailed outline} In Section~\ref{sec:data-concepts} we present basic definitions, data structures, and notations used all along this paper. In Section~\ref{sec:related-works}, we present related works from the domains of secret sharing and all-or-nothing encryption as well as point out their limitations. 
In Section~\ref{sec:model}, we introduce a new security model (denoted as SAKE) and show that, under our storage site-by-storage site threat model, it is as secure as the more constraining one introduced recently in \cite{bib:bastion}. We finally highlight the main technical challenges of one-round encryption under key exposure, which to our knowledge we are the first ones to properly address. In Section~\ref{sec:ssake}, we illustrate the relevance of our approach by describing a new scheme (denoted as $\caon$) providing protection against key exposure that reduces the complexity of the post-encryption processing by half with regard to the state-of-the-art fastest scheme. In Section~\ref{sec:analysis}, we do not only present its security proof but also revisit the security analysis of a recent relevant scheme. As an alternative to our $\caon$ scheme, in Section~\ref{sec:analysis:keccak}, we introduce a scheme (denoted as $\kecchare$) that totally gets rid off the additional post-processing overhead, and describe both an instantiation using blockciphers, and also implement it with Keccak. Before concluding in Section~\ref{sec:conc} , Section~\ref{sec:comparison} focuses on comparisons between our two new proposals and the state-of-the art techniques in terms of complexity, memory occupation, and verification of security properties. Complexity results are confirmed by performance evaluation.

%% file: data_structures_and_notations.tex
\section{Basic Definitions and Notations}
\label{sec:data-concepts}

We identify the following definitions and basic data structures that mostly correspond to classical concepts concerning block cipher encryption and computational secret sharing scheme. Our main unit of length is a \emph{block}, which is a sequence of bits of size $\blen$, where $\blen$ is our \emph{security parameter}. All secret keys $\key$, headers, initialization vectors $\iv$ etc. will be of length one block, such as the inputs/outputs $\pbl_i$/$\cbl_i$ of the blockciphers $\bc$  that we will consider%
\footnote{Notwithstanding, our blockcipher instantiation of $\kecchare$ will use a blockcipher with \emph{two-blocks-long key} $\mathcal{L}:=(\key||\iv)$, where $\key$ is the actual one-block-long key of our final scheme. But the blockcipher itself will still process with one-block long inputs and outputs $\pbl_i$ and $\cbl_i$: e.g. AES256, which processes 128 bits blocks.}.
Thus, a block is necessarily \emph{small}, compared to shares (see \cite{bib:Dworkin} and below) which can be \emph{much bigger} (we focus on large data protection as perfect secret sharing instead of symmetric encryption can be applied for small data, solving the problem of key exposure). We note $+$ or $\oplus$ the XOR operation between two blocks, which is the sum of binary vectors in $\{0,1\}^\blen=\F_2^\blen$. We will sometimes add a block with a number $i$ (typically the counter of CTR mode). By this abuse of notation, we simply mean the binary writing of the number $i$, seen as a vector in $\F_2^\blen$.

\begin{itemize}

\item \textbf{Plaintext $\plain$}: the data to be protected, $\plain$ is a sequence of $\plainl$ bits and (in the context of block cipher encryption) of blocklength $\clen-1$ blocks $\pbl_i$, where $i$ vary from $1$ to $c-1$.
\footnote{In \cite{bib:bastion}, indices are from $1$ to $m$ ($c=m+1$).}

\item \textbf{Secret $K$, $IV$, and $r$:} We distinguish three secret values. A secret key $\key$ of size $\keys:=\blen$ of one block, which will be never stored within the data and given only to the persons entitled to read the data (this will be abstracted out in our model). By contrast, the two other secret quantities $r$ or $\iv$ (corresponding to a pseudorandom header/nonce or to an initialization vector) are stored in the storage sites, but in a secret-shared manner: they are "mixed within the shares".

\item \textbf{Ciphertext $\ciph=\enc(\plain,\key)=(C_i)_i$}: plaintext encrypted with a randomized keyed encryption scheme $\enc$. $\ciph$ has a total bitsize of $\ciphl$ bits. If encryption was done using a blockcipher, then $\ciph$ is a sequence of $\clen$ blocks. Indices of the $C_i$ vary from $0$ to $c-1$. \footnote{In \cite{bib:bastion}, they are noted $y'_1$ to $y'_{m+1=n}$, where $n$ denotes the number of blocks.}

\item \textbf{Transformed ciphertext $\ciph'=(C'_i)_i$}: in case when a linear transform is applied to the ciphertext $\ciph$, the $\ciph'$ notation is used to distinguish between the ciphertext and the ciphertext after the transformation. \footnote{In our $\caon$ scheme, indices of the $C'_i$ vary from $1$ to $c-1$. Whereas in \cite{bib:bastion} they are noted $y_1$ to $y_{m+1=n}$. In our $\kecchare$ scheme the transformed ciphertext $\ciph'$ just consists in the ciphertext $\ciph$ where the initialization vector has been removed. Moreover, when compared to \cite{bib:bastion}, we reverted the primes between the $\ciph$ and the transformed $\ciph'$; our $\ciph'$ is one block smaller; and our $\ciph$ are of length $c$ instead of $\nbs$ (the letter $\nbs$ being reserved for the number of shares).}

\item \textbf{Fragment $F_i$ for $i=1\dots,\nbs$}: we partition the transformed ciphertext into fragments $F_i$. For simplicity, if the transformed ciphertext is composed of $\clen-1$ blocks (or $|C|$ bits), then each $F_i$ is a group of $\frac{\clen-1}{\nbs}$ consecutive blocks (or $\frac{\ciphl}{\nbs}$ bits). We assume for convenience that $\clen-1$ and $\ciphl$ are divisible by the number of shares $\nbs$.
%$F_i=C'_{\frac{\clen-1}{n}\times(i-1)},\ldots,\,C'_{\frac{\clen-1}{n}\times(i)-1}$. We assume for convenience that $\clen-1$ is divisible by $\nbs$.

\item \textbf{Share $\s_i$ for $i=1\dots,\nbs$}: the shares are by definition all, and only all, what is stored in the storage sites. They typically consist in the above fragments, plus a share of some secret quantity $\iv$.  In any case, the shares never contain (to be precised) the secret key $\key$. \footnote{In \cite{bib:bastion} shares are equal to the fragments.}

\item \textbf{"All-or-nothing transform" $\trans(\key,\plain)$:} randomized transformation mapping the plaintext $\plain$ into a set of $\nbs$ shares $\s_1,\ldots,\s_{\nbs}$. It possibly uses a secret key $\key$ as additional input. It also always uses a random input $\iv$, that we do not make explicit in the input variables. It uses a public symmetric encryption scheme $\enc$ as an underlying mechanism. In our new security model, "all-or-nothing" should be understood as "all storage sites/shares". 

\item \textbf{Random oracle (RO)}: a function $\co:\bigr(\F_2^{2\blen}\bigl)\lra \bigl(\F_2^\blen\bigr)^\infty $ of arbitrarily long output, and taking as input small seeds (for example two blocks: $\co(\key||\iv)$). We note $\co_{\clen-1}$ for the oracle with output truncated to $\clen-1$ blocks. A key property of random oracles that we will use is that, when fixing the first input block $\key$, the function $\co(\key||\ast)$ is \emph{still} a random oracle, this time of one-block input, moreover \emph{independent} from all the other oracles $\co(\key'||\ast)$.

\item \textbf{Ideal keyed blockcipher $\bc$}: a symmetric encryption function that maps blocks to blocks.

\end{itemize}

\noindent \textbf{Detailed definition of a random oracle $\co$} We refer to the definition presented in \cite[§13]{katz}. When queried on a new input, $\co$ outputs a string of independent uniformly distributed bits, and when queried on a previously queried input, $\co$ returns the same output. We will use the fact that a RO satisfies in particular the strictly weaker property of being a pseudorandom generator ---also known as (idealized) seeded pseudorandom number generator (PRNG) \cite[Definition 3.15]{katz}. See \ref{sub:nail} for a discussion of why the random-looking property of the output of a pseudorandom generator is in general \emph{not} preserved when part of the seed is leaked, while the random-looking property of the output of a random oracle $\co$ \emph{is} preserved even if part of the seed is leaked. This is the main technical subtlety of this paper, and the reason why we pay much care in security analyses under key exposure.

\noindent \textbf{Keyed pseudorandom function/permutation (PRF/PRP)} We refer to \cite[Definitions 3.24 \& 3.26]{katz}. A PRF $f$ is a function: $f:\F_2^{\blen}\times \F_2^{\blen} \lra\F_2^{\blen}$, such that when the first input component $\key$ is chosen at random and fixed once for all, then the output of $f(K,\iv)$ on several randomly chosen secret $\iv$ is indistinguishable from independent uniform random blocks. % A PRP for our security proofs when the key is \emph{leaked}, we will need the strictly stronger notion of an ideal blockcipher

\noindent \textbf{Definitions of ideal keyed blockcipher and ideal random permutation} Shannon's perfect blockcipher model is an oracle $\bc:\F_2^\blen\times \F_2^\blen\rightarrow\F_2^\blen$ such that, on every query with a new key $\key$ (the first input component), it outputs a fresh new random permutation oracle $\bc_\key: \F_2^\blen\ra\F_2^\blen$ \cite{blackidealbc}. In turn, let us recall informally that a random permutation oracle (as assumed by the random permutation model: RPM) is a tuple of functions  $\perm,\perm^{-1}:\F_2^\blen \ra\F_2^\blen$ operating on blocks and inverse of each other. They have the following ideal behavior: on every input to $\perm$ $\bigl\{$not queried to $\perm$ nor output by $\perm^{-1}$ before$\bigr\}$, then $\perm$ outputs a value chosen uniformly at random among all values $\bigl\{$not output by $\perm$ nor input to $\perm^{-1}$ before$\bigr\}$; and the inverse oracle $\perm^{-1}$ behaves accordingly in the other direction. These three ideal primitives (ideal BC, RO, and RPM) were recently proven to be equivalent to each other.
\footnote{Ideal BC is efficiently instantiable from RPM \cite{lampe} and also from RO \cite{holenstein}. Finally, a RO with \emph{arbitrary long variable input and output} can itself be instantiated from a sponge function calling an ideal random permutation \cite{keccak}.}

%\footnote{Ideal BC is efficiently instantiable from RPM: see Lampe et Seurin Asiacrypt 2013 \cite{rpmbc}, and also from the random oracle (RO) model: see Holenstein et al STOC 2011 \cite{holbcro} (fixing Coron-Seurin Crypto 2008 \cite{coronrobc}). Finally, a RO with \emph{arbitrary long variable input and output} can itself be instantiated from a sponge function calling an ideal random permutation (RPM): see \cite{keccak}.}

\noindent \textbf{Chosen plaintexts indistinguishability ind1 and ind-CPA, po\-lynomial adversary and negligible functions} We consider an adversary $\A$ who can make a total number of queries which we note $\quer$. Let us define a \emph{negligible function} as any fixed polynomial in $\blen$, $\clen$, and $\quer$ divided by $2^{\blen}$. Implicitely, this means that our adversary is \emph{polynomial}. Thus what we will call a "negligible event" should be understood as something which occurs with probability equal to a negligible function. Likewise, when we say that the adversary has "negligible advantage" in a game, we mean that she wins it $50\%$ of the time, plus or minus a negligible percentage. With respect to these conventions, an encryption scheme with plaintexts indistinguishability under queries with \emph{one-shot keys} ---a.k.a. ind1--- is defined in \cite[Definition 3.9]{katz}. Indistinguishability under multiple queries with the \emph{same key}, which we simply note "ind-CPA", is defined in \cite[Definition 3.22]{katz}.

%% file: relevant_work.tex
\section{Related Work}
\label{sec:related-works}

This section presents an overview of methods reinforcing confidentiality by dispersing data over different storage sites. 

%\label{sub:pss}  \label{sub:css} \label{sub:aon}
\textbf{Perfect secret sharing (PSS)}   Perfect secret sharing~\cite{bib:blakley,bib:shamir} with threshold $\nbst-1$ transforms data into a set of $\nbs$ shares, $\nbst$ of which are needed for data reconstruction. Strictly less than $\nbst$ shares provide no information \textit{whatsoever} about the initial data, so the information is protected against adversaries unable to collect the required threshold of $\nbst$ shares. Shamir's perfect secret sharing scheme (PSS) \cite{bib:blakley,bib:shamir} is information theoretically secure. This unconditional security comes at the cost of a $\nbs$-fold increase in the volume of the data to be stored as \emph{each of the shares is of the size of the data itself}. 
 
\textbf{Computational secret sharing (CSS)}
In it's vanilla version, Krawczyk's Secret Sharing Made Short~\cite{ssms} consists in encrypting the plaintext, then dividing the ciphertext into $\nbs$ fragments. Each storage site receives a "share", consisting in one fragment plus a share of the encryption key under a PSS with threshold $\nbst-1$. In \cite[Theorem 3]{ssms} this scheme is proven to be a secret sharing scheme with threshold $\nbs-1$, under computational assumptions. As we detail later in §\ref{sub:krawbreak}, \emph{Krawczyk's CSS breaks down under key exposure}.
 
\textbf{All-or-nothing encryption (AON)} 
An all-or-nothing (AON) encryption makes the ciphertext decryptable only when complete. This security property can be achieved using a pre- or post-processing of encrypted data denoted as an all-or-nothing transform (AONT). A detailed overview of AON schemes can be found in \cite{QKLQM19}, and a recent one is \cite{peaont}.

For space constraint we defer to Appendix \ref{sub:aonts} the detail of Rivest's AONT scheme,  Boyko \& Desai's AONT schemes and formalization of what an AONT is. Although the definitions differ, one can roughly understand it is an efficiently invertible transformation such that, given two chosen plaintexts, an adversary knowing all but one blocks of their transforms, cannot distinguish between them. The transformation is composed of two encryption rounds using two different keys, only one of them will be exposed.

%Rivest's AON scheme consists in an AONT, which uses in practice one encryption round, followed by a second encryption, not to mention one hash computation of a ciphertext in the middle. Performance is thus an issue. Desai~\cite{desai} proposed an alteration of the Rivest's scheme by replacing the hash of the ciphertext with a sum of all the ciphertext blocks. However, it still requires two encryption rounds. Notice that a modification of Rivest's sole AONT transform, called AONT-RS~\cite{resch}, is applied inside the IBM Cloud Object Storage~\footnote{https://www.ibm.com/cloud/object-storage. Formerly: Cleversafe.}. See \cite{martin} for a structural security flaw in "AONT-RS".

%\subsubsection{Stinson's linear all-or-nothing transforms and the Bastion scheme}
%\label{sub:stinson}

Stinson~\cite{stinson,stinsonnew} formalized a linear all-or-nothing transform as an invertible matrix $Mat$ of size $\nbs\times\nbs$,  that maps an "unknown" input vector $\x$ of length $\xlen$ to a "partially known" output vector $\y$, such that for each index $i$, and every missing coordinate $j$%, we have: $$\mathcal{H}(x_i|y_1,\dots,y_{j-1},y_{j+1},\dots,y_\nbs)= \mathcal{H}(x_i) $$ where $\mathcal{H}$ is the entropy. That is, for every unknown coordinate $x_i$
, an attacker who manages to learn all coordinates of $\y$ but $y_j$ will learn no more information on $x_i$ than what she knew a priori. However, we would like to emphasize that the standalone property of a linear AONT is not enough to obtain a cryptosystem. Indeed, as emphasized by Boyko in \cite{boyko}%
~\footnote{"However, the linear constructions of Stinson would definitely not be secure in that model, since it is easy to come up with linear relations among the elements of $\x$ by looking at just a few elements of $\varphi(\x)$ (in fact, since $\varphi(\x)$ is linear and deterministic, every output of $\varphi(\x)$ gives a linear relation on elements of $\x$."}%
, an attacker may still learn \emph{correlations} between unknown coordinates $x_i$, potentially ruining the ind-CPA security property of the ciphertext. %In a usual use case, the input vector $\x$ corresponds to the ciphertext $\ciph$ and the output vector $\y$ to the transformed ciphertext $\ciph'$, which is then fragmented into shares that are dispersed over several storage sites. As $\ciph$ cannot be then reconstructed from an incomplete $\ciph'$, data is protected against attackers unable to gather all the $\ciph'$ shares.

Recently introduced, Bastion scheme~\cite{bib:bastion} builds on a variant of Stinson's linear AONT. The plaintext is first encrypted, then the ciphertext is transformed using a square matrix $Mat^{Bastion}$ such that:  (i) all diagonal elements are set to 0, and (ii) remaining off-diagonal elements are set to 1. 

As a result, each block of the ciphertext is XOR-ed with all other ciphertext blocks. The transformed ciphertext is then claimed to be protected against key exposure and all but two ciphertext blocks exposed (one more block than in the case of Stinson's AONT). The advantage of Bastion's approach is that they require only a single encryption round and thus is much faster than the Rivest's initial proposal (see Figure \ref{fig:performance}). 

%% file: security_model.tex
\section{Encryption under Key Exposure: Why a New Model?} 
\label{sec:model}

We motivate in §\ref{sub:sake-motivations} our alteration of existing security models in order to address resiliency of shared data under key exposure, that we present in §\ref{sub:defsake}: $\nbs$-Shares Access under Key Exposure, that takes as parameter $\nbs$ the number of shares (=storage sites) instead of the amount of compromised blocks. Then in §\ref{sub:krawbreak} we explain why previous schemes, such as Krawczyk's SSMS, break down under this scenario of key exposure. We give $\nbs$-$\sake$ a more formal definition in \ref{sub:defformalsake}, that we relate to the classical ind-CPA security. In §\ref{remark-on-r}, we give a soft overview of the challenges to overcome to achieve efficient $\nbs$-SAKE schemes, and discuss previous schemes achieving SAKE security. In \ref{sub:sake-cake} we compare SAKE to CAKE, and emphasize that they are \emph{equivalent} against an adversary corrupting all but one storage sites. Finally in \ref{sub:nail} we nail down the main technical subtlety that will make the difference between encryptions which are directly SAKE, and those which are not (and thus which will require extra post-processing).

\subsection{Motivations} \label{sub:sake-motivations}

The state-of-the-art security model relevant for evaluating the security of a computational secret sharing scheme under key exposure was introduced in \cite{bib:bastion} and is denoted as Ciphertext Access under Key Exposure (CAKE). A scheme is denoted to be $(\clen-\lambda)$ CAKE secure when it resists an attacker able to access \emph{any} $(\clen-\lambda)$ blocks of the dispersed ciphertext as well as the encryption key. This model, inspired by all-or-nothing encryption, still does not take into account the fact that, once transformed, the ciphertext will be cut into $\nbs$ shares containing a fraction of e.g. $\frac{\clen}{\nbs}$ blocks each%
\footnote{We have different notations than \cite[§3.1]{bib:bastion}.}.
These shares will be then distributed over $\nbs$ different storage sites or servers, preferably belonging to independent storage providers.  Thus, the power given to the adversary to obtain \emph{any} ciphertext blocks that she wants may be overkill in this context. The actual difficulty for an attacker lies in looking for and acquiring the totality of these dispersed shares, which is equivalent to compromising all the storage sites. Indeed, once an attacker compromises a storage site, she will be most probably able to acquire all the blocks of the share $S_i$ stored at this site ~\footnote{A rare exception to this scenario would be in the case of a memory leak attack.}. 

Motivated by the observation described above, we propose a relaxed security model, in which \textit{we want to protect the data shares against an attacker able to compromise all but one of the storage sites}. This alteration in perspective allows us to design faster secret sharing schemes without in practice compromising on security.

\subsection{Shares Access under Key Exposure: $n$-SAKE} \label{sub:defsake}
\label{sub:sake-description}

%Let us define our security model, and compare it with the existing $(\clen-\lambda)$-CAKE notion introduced previously in \cite{bib:bastion}.

We introduce now our threat model that addresses a scenario where data $\plain$ is transformed into $\nbs$ shares $\trans(\key,\plain)= \s_i$, $i=1,\ldots,\nbs$ and these shares are then dispersed over $\nbs$ storage sites. In this model, we can have two types of attackers. The first type of attackers \textbf{(1)} will be able to compromise all the storage sites and gather all the shares but will not be able to access the symmetric encryption key $\key$. They are typically honest-but-curious cloud providers who would have access to all the shares. The second type of attacker \textbf{(2)} will be able to obtain the key, e.g. due to bad key management, but will not be able to compromise the totality of the storage sites. 

We now formulate two security properties that characterize a scheme resisting the presented threat model:

\begin{enumerate}
\item \textbf{Security property (1) - classical ind-CPA property}: In the scenario where the adversary possesses all the shares but has no information about the key, we ask for the classical indistinguishability under Chosen Plaintext Attack (ind-CPA) as defined for instance in \cite[p. 74]{katz}. 

\item \textbf{Security property (2) - ind-CPA property of all-but-one shares under key exposure}: 
In the scenario where the attacker could get the key and get to all but one of the storage sites, we ask for the indistinguishability under chosen plaintext attack and under key exposure of the $\nbs-1$ shares in possession of the attacker.
\end{enumerate}

\subsection{Comments on the $n$-SAKE definition} \label{sub:krawbreak}

Property \textbf{(1)} is easy to match since it suffices to encrypt data using any 
scheme having the ind-CPA property before transforming them into shares and, most importantly , \textit{not to hide the encryption key within the shares}. Indeed, if the key is secret shared and the key shares are attached or appended to the data shares, as in the case of Krawczyk's SSMS in its original form, an attacker getting the shares will automatically get the encryption key. Similarly, in the case of Rivest's or Desai's AONT, the key can be recovered once the ciphertext is complete. So in all these schemes \textbf{(1)} doesn't hold.

By contrast, property \textbf{(2)} will be much harder to achieve. Let us give a taste of why. Let us consider the Krawczyk's CSS scheme where the data owner \emph{doesn't} "mix" at all the encryption key within the shares, but instead keeps it for himself (and for the other persons accredited). Let us note $\enc$ the baseline encryption scheme used in Krawczyk's and $\key$ the secret key. Let us suppose informally that $\enc$ operates as a stream cipher, so that, given the key, the first blocks of the ciphertext can be decrypted into the first blocks of the plaintext. Now consider an adversary who manages to obtain this key $\key$, and who chooses two plaintexts of same length which differ in their initial block. Then, this adversary can distinguish between these plaintexts as soon as she is given the share containing the initial blocks of the ciphertexts.  

Notice that we guarantee nothing against an adversary who has access to all the shares and to the decryption key $\key$. On the contrary, we ask for the plaintext to be \emph{efficiently reconstructible} from these two ingredients. Additional means of protecting the plaintext are outside the scope of this paper.

\subsection{Formal definition of $n$-SAKE security} \label{sub:defformalsake}
We consider a \emph{public randomized keyed invertible transformation} "$\trans$" that maps a plaintext $\plain$ to $\nbs$ bit strings, that we call "shares":
\begin{equation}\trans(\key,\plain) \longrightarrow \s_1,\dots,\s_\nbs.\end{equation}
 
\noindent and such that the inverse is efficiently computable given all the shares and the key $\key$. We don't explicitly note here the random component of the input: $\iv\in \{0,1\}^\blen$ of $\trans$ used during the transform (e.g. the Initialization Vector in block cipher encryption, or the sponge header \cite{duplex} - at this point, we do not specify the encryption details). We say that $\trans(\ast,\ast)$ is a $\nbs$-$\sake$ transformation if and only if:

\textbf{(1)} The following encryption scheme is ind-CPA secure: select once for all a secret key $\key$ at random, then on every plaintext $\plain$, output all the shares $\trans(\key,\plain)$.

\textbf{(2)} We now give a formal definition of the $n$-SAKE property \textbf{(2)} based on the classical ind-CPA game. Let us fix an encryption key $\key$ that will be used by the "$\trans$" transform and that will be supposed from now on known to the adversary (as we want to deal with the case of a key leakage). 
 
The polynomial adversary $\A$ has access at any time ($\A$ is said  "adaptive") to an oracle which performs the encryption scheme $\enc$ applied inside of "$\trans$" (and its inverse $\enc^{-1}$) on any (or ciphertext) of $\A$'s choice. It also performs $\trans$ (or its inverse) on any plaintext (or set of $\nbs$ shares) of $\A$'s choice.

Here is the $n$-SAKE security game \textbf{(2)} between $\A$ and $\co$, that defines the ind-CPA security in the context of key exposure.

\begin{enumerate}
    \item $\mathcal{A}$ outputs to $\co$ a pair of plaintexts $\plain^{0}$ and $\plain^{1}$ of the same length.
    \item $\co$ chooses randomly a bit $b\in {0,1}$. 
    \item $\co$ performs the randomized transformation $\trans(\key,\plain^b)$ that gives $\nbs$ shares $\s^{(b)}_1,\ldots,\s^{(b)}_{\nbs}$.
    \item $\mathcal{A}$ adaptively queries $\nbs-1$ indices of shares and $\co$ gives them to her.
    \item $\A$ outputs to $\co$ a bit $b'$.
    \item The output of the experiment is defined to be $1$ if $b=b'$ and $0$ otherwise. In the former case, we say that $\mathcal{A}$ succeeds.
\end{enumerate}

%Similarly to the ind-CPA game, an AON scheme is not $(\nbs-\lambda')$ shares - CAKE secure if the adversary has a non-negligible chance of winning the presented game.
If the adversary has no non-negligible advantage in this game, with respect to the length of one block $\blen$ as security parameter, then we say that the transformation $\trans$ fulfills the property \textbf{(2)} of $\nbs$-SAKE.

A careful reader will notice that if shares are badly chosen in some scheme, then this scheme \emph{cannot} satisfy our \textbf{(2)} security requirement of $\nbs$-$\sake$ (e.g. if a share is always empty, or two shares always equal, then an adversary can recover everything from $\nbs-1$ shares). In this paper we care about designing schemes that \emph{are} $\nbs$-$\sake$, not schemes that are not.

\subsection{How to design more efficient schemes?} \label{remark-on-r} 

As regards the random parameter $\iv$, although it is classically put as a header of the ciphertext in classical ind-CPA encryption schemes, we just stressed in \ref{sub:krawbreak} that the security property \textbf{(2)} completely breaks down if the adversary recovers this random parameter. Indeed, she could then decipher the beginning of the ciphertext with this $\iv$. So in particular, we have to hide it in a manner that any $\nbs-1$ shares give no information about it.

Note that all previous all-or-nothing schemes are based on the same pattern: perform a first transform "AONT", which is an encryption where the encryption key is hidden within the ciphertext (in a more size-efficient way than in Krawczyk's SSMS). Then encrypt a second time, this time with a key $\key$ that is kept by the user. In particular, Rivest's, Boyko's and Desai's AON verify the $\nbs$-$\sake$ property with respect to the second key. In the Bastion \cite{bib:bastion} scheme, as in our schemes, the Initialization Vector/random nonce $\iv$ used during the encryption plays the role of the "first key" $\iv$ that is hidden within the ciphertext (Bastion) or shares (us). The real challenge in designing efficient schemes matching this double security objective \textbf{(1)} and \textbf{(2)}, is to prevent attacks such as in §\ref{sub:krawbreak}. But just hiding the $\iv$ within the shares is not enough, since parts of ciphertexts can be distinguished if the encryption has bad randomness properties under key exposure: see §\ref{sec:revisiting-assumptions}. This is why we will still need a (small) post-processing of the ciphertext in $\caon$.

\subsection{Difference with the CAKE model}
\label{sub:sake-cake}

We point out here some remarks about the differences between the CAKE definition from \cite{bib:bastion} and SAKE security properties. We discuss as well different possibilities concerning the size of the shares and their impact on the security of a secret sharing scheme.

The SAKE security property is clearly inspired by the $(\clen-\lambda)$ CAKE property. We formulate therefore the two following correspondences between the two properties:

\begin{remark}{\textbf{From CAKE to SAKE:}} 
\label{remark:CAKE-SAKE}
To make it simple: consider a $(\clen-2)$-CAKE scheme, such as the one of \cite{bib:bastion}, which outputs a ciphertext of length $\clen$. Then in the highly typical cases where $\nbs\leq 2\clen$, splitting this ciphertext into $\nbs$ shares of size at least two blocks each yields a $\nbs$-SAKE scheme%
\footnote{In full generality: a $(\clen-\lambda)$ CAKE secure scheme is SAKE secure if the sum of blocks in any subsets of its $\nbs-1$ shares does not exceed $(\clen-\lambda)$.}.
\end{remark}

\begin{remark}{\textbf{From SAKE to CAKE:}} 
\label{remark:SAKE-CAKE}
A $n$-SAKE secure scheme is $(\clen-\lambda')$ CAKE secure where $\lambda'$ denotes the minimum of all possible sums of combinations of $\nbs-1$ of the shares. To make it simple: in the highly typical cases where $n\leq \clen$, then a $n$-SAKE scheme is only $\nbs-1$-CAKE. Indeed, if a CAKE adversary chooses $\nbs$ ciphertext blocks in $\nbs$ distinct shares, then SAKE doesn't guarantee anything.
\end{remark}

For the sake of simplicity, we will define all shares in our schemes as having the same size. This definition could be modified in order to fit a non-uniform distribution of ciphertext among the shares. This would include cases like having two shares where one share contains $\clen-1$ blocks of the ciphertext and the other just one block (that could be useful for instance when wanting to have a large outsourced fragment of data in the cloud and one small fragment kept at the user's device).

In a multi-cloud storage scenario $\nbs$ is rather small \cite{bib:depsky} due to the burden coming with the subscription to a new storage provider. In a case where data is dispersed over multiple servers, $\nbs$ is rarely greater than 20-30 \cite{bib:aont-rs}.

\subsection{Nailing down the main technical difficulties} \label{sub:nail}
 Let us consider a keyed permutation $f(\ast,\ast)$, which is at least a PRP as in §\ref{sec:data-concepts} (or refer to \cite[Definition 3.24]{katz}), or even which satisfies the stronger notion of an ideal blockcipher. Then consider the following classical "counter-mode" strings generator: fix a secret key $\key$ once for all, then on every query, sample $\iv\in\F_2^\blen$ at random then output:
  \begin{equation} \label{eq:prng} \bigl[ f(\key,\iv),f(\key,\iv+1),f(\key,\iv+2),\dots\bigr] \; .\end{equation}
 These outputs are indistinguishable from uniform independent random bit strings. Said otherwise, construction \eqref{eq:prng} is a pseudorandom generator in the sense of \cite[Definition 3.24]{katz}.
  
On the other hand, if the $\key$ is leaked to the adversary, then \emph{this is false} (see Appendix \ref{sub:cexctr} for a detailed explanation). In §\ref{sec:revisiting-assumptions} we explain how a previous scheme felt into this safety trap along its proof under key exposure. Notice that a similar breakdown occurs if \eqref{eq:prng} is replaced by the classical "chained-based" (CBC) mode of operation and, worse, this is then the case even if $f$ is just a PRF (not a PRP anymore).

We will fix this problem with post-processing of the output in §\ref{sec:ssake}-§\ref{sec:analysis}, whereas in §\ref{sec:analysis:keccak} we will fix it with the following trick: \emph{if we replace the construction} \eqref{eq:prng} using a random oracle $\co$ as follows: $\iv\ra\co(K||\iv)$, then \emph{it is true} that this string generator remains pseudorandom even when $\key$ is leaked to the adversary. We will use this crucially in the proof of Main Theorem \ref{th:romain}. Interestingly, we show in \ref{sub:roaes} that the construction \eqref{eq:prng} \emph{switched upside down}  [$\iv$ becomes the secret key and $\key$ the variable random input] is secure under leakage of the secret $\iv$, when assuming an ideal random behavior (ideal blockcipher model) on $f(\ast,\ast)$.

%% file: ssake_description.tex
\section{SSAKE: fast block-cipher based CSS protecting against key exposure}
\label{sec:ssake}

% A key exposure resilient scheme, with twice less overhead on top of encryption and based on counter-mode of encryption

We introduce a new computational secret sharing scheme denoted as $\caon$ : \textbf{S}ecret \textbf{S}haring \textbf{A}gainst \textbf{K}ey \textbf{E}xposure. In the next section we will prove that it satisfies the previously introduced $SAKE$ security notion:

\begin{main} \label{th:mainssake} In the ideal blockcipher model, the scheme $\caon$ satisfies $n$-$\sake$ security with respect to the key parameter $\key$ used for the blockcipher encryption. \end{main}

The advantage of this scheme is that it is \emph{versatile} because it uses a standard blockcipher in the CTR mode as the encryption scheme inside the transformation $\caon$ (for instance the standard AES), so that the transformation can be applied on already encrypted data. The transformation $\caon$ is composed of four steps: 
\begin{enumerate}
    \item Encryption of the plaintext into a ciphertext using a blockcipher (like AES) in the CTR mode.
    \item Transformation of the ciphertext using a linear transform creating dependencies between the blocks.
    \item Splitting the transformed ciphertext into $\nbs$ fragments $F_i$ composed of consecutive blocks.
    \item Applying a perfect secret sharing scheme to the initialization vector $\iv$ used during the encryption to produce $\nbs$ shares $\iv_i$, and attaching these shares to the fragments: an output share $S_i$ is then the concatenation of the transformed ciphertext fragments $F_i$ and the IV's share $\iv_i$.
\end{enumerate}

We consider a keyed blockcipher $\bc$, which is a publicly known keyed function operating on blocks of size $\blen$ bits and which outputs $\blen$ bits (when designing the scheme we think of the most common symmetric encryption block cipher AES with blocks of 128 bits). Here, the secret key of the blockcipher is of size $\keys=\blen$ one block.

The $\caon$ algorithm starts with plaintext $\plain=P_1,\ldots,P_\plen$ encryption using the $\bc$ in Counter Mode $CTR$ (we chose this mode instead of CBC for parallelizability, and simplicity of Proposition \ref{hypodiff}). This results in a  ciphertext $\ciph=C_0,\ldots,C_{\clen-1}$, composed of $\clen$ blocks. For the sake of completeness, let us just remind that such encryption consists to one-time pad the vector $[0||\plain]\in (\F_2^{\blen})^{\clen}$ ($\plain$ concatenated with zero block appended) with the following vector (which is \emph{not} pseudorandom: see the explanation Section \ref{remark-on-r}, as well as §\ref{sec:revisiting-assumptions} and Appendix \ref{sub:cexctr}):

     \begin{equation}[\bc(\iv),\bc(\iv+1),\dots,\bc(\iv+\clen-1)]\in (F_2^{\blen})^{\clen}\end{equation} 
      generated from a "seed" block $\iv\in\F_2^\blen$ sampled at random. The output of this operation is the ciphertext 
      \footnote{We ask for the first block of ciphertext to be $C_0=\bc(\iv)$, in order to be in the setting of Hypothesis \ref{hypodiff}. We could instead have stuck with the classical choice $C_0:=\iv$, and modified Proposition \ref{hypodiff} accordingly.}
      \begin{equation}\ciph=[C_0,\dots,C_{\clen-1}]\in (\F_2^{\blen})^{\clen} \end{equation}
      of $\clen=(\plen)+1$ blocks.
    
A \emph{noninvertible} linear transformation is then applied to $\ciph$ transforming it into $\ciph'=[C'_1,\dots,C'_{\clen-1}]$ of length $\clen-1$ blocks as follows: $C'_i:= C_i + C_{i-1}$ $\forall i\in[\clen-1]$.
\emph{Additionally}, the initialization vector block $\iv$ is split using a perfect secret sharing scheme (PSS) with adversary threshold $\nbs-1$ into $\nbs$ shares $\iv_1,\ldots,\iv_\nbs$: for example, an additive secret sharing. So the original ciphertext $\ciph$ can be recovered from both $\ciph'$ and the shared $\iv$.
The transformation can be seen as the right matrix multiplication $\ciph'=\ciph\cdot Mat^{SSAKE}$, where:
    %$$=Mat^{SSAKE}\times(\ciph) ,\text{where:} $$
    \begin{equation}Mat^{SSAKE} = \begin{pmatrix}[cccccc]%\ciph':=\ciph\cdot \begin{pmatrix}{ccccccc}
 1 & 0 & 0 & 0& . & 0 \\
 1 & 1 & 0 & 0& . & 0 \\
 0 & 1 & 1 & 0& . & 0 \\
 0 & 0 & 1 & 1& . & 0 \\
 0 & 0 & 0 & 1& . & 0 \\
 . & . & . & .& . & . \\
 0 & 0 & 0 & 0& . & 1 \\
\end{pmatrix}\end{equation}

Transformed ciphertext $\ciph'$ is split into $\nbs$ fragments. This can be done in various ways. The simplest way is to just create the fragments from large chunks of consecutive $\frac{\clen-1}{\nbs}$ blocks. Consecutive blocks can be also dispersed over different fragments - this would reveal less information to an attacker that somehow managed to obtain the $IV$ but has no knowledge about the plaintext.

    \begin{figure}[htbp]
\begin{algorithmic}[1]
%\Function{SSAKE}{$\plain,\nbs$}
\Function{SSAKE}{K,PLAIN}
    \State Encrypt $\plain$ into $\ciph=[C_0,\ldots,C_{\clen-1}]$ of $c$ blocks using blockcipher $\bc$ with key $\key$ in counter-mode.
    \State Linear transform:
    \For {each $i=1,\ldots,\clen-1$}  
	\State Compute $C'_i=C_{i-1}+C_i$
    \EndFor
    \State Fragment $\ciph':=[C'_1,\ldots,C'_{\clen-1}]$ into $\nbs$ fragments $F_1,\ldots,F_{\nbs}$ 
    \State PSS $\iv$ into shares: $PSS(\iv)= \iv_1,\ldots,\iv_\nbs$
    \For {each $i=1,\ldots,\nbs$}  
    \State Output the concatenation $S_i:=\iv_i||F_i$. 
    \EndFor
    %\State First block (version 2): $C_0$ is secret shared into $k$ blocks that are attached to the final shares.
\EndFunction
\end{algorithmic}
\caption{\textit{Pseudo-code of the $\caon$ transformation for $\nbs$ shares, using a secret key $\key$ generated at random once for all.}}
\label{fig:transform}
\end{figure}

For reconstruction: recover $\iv$ from all the shares, then, for someone who \emph{knows} the key $\key$, deduce $C_0=\bc_K(\iv)$, then deduce the $C_i$ sequentially, then decipher them: $C_i-\bc_K(\iv+i)$.

%% file: security_analysis.tex
\section{Security Analysis of the $\caon$ scheme (and of previous work)} \label{sec:analysis}
We first prove in §\ref{sub:proofcaon} that $\caon$ achieves $\nbs$-SAKE security, then in §\ref{sec:revisiting-assumptions} we revisit the security proofs of \cite{bib:bastion} and highlight ---one more time--- the technical issues raised by the setting of key exposure and finally single out in §\ref{sub:fixbastion} a sufficient hypothesis under which the scheme of \cite{bib:bastion} is secure.

\subsection{Proof of Main Theorem \ref{th:mainssake}} \label{sub:proofcaon}

(1) $\caon$ is trivially ind-CPA for an adversary who ignores the secret key $\key$ and is given all the shares. Indeed, the shares are the result of a public transformation (the linear transform-then-sharing) whose sole input is the ciphertext $\ciph$. But this ciphertext $\ciph$ itself is the plaintext encrypted with the classical counter-mode blockcipher encryption scheme with secret key $\key$. Thus any two ciphertexts $\ciph$, and thus any two transforms of them with a public transformation, are well-known to be indistinguishable under chosen plaintext attacks (see e.g. \cite[Theorem 3.30]{katz}).

The proof of (2) is based on the following indistinguishability property:
 
\subsubsection{Main technical result}
We place ourselves in the ideal permutation model (RPM) recalled in §\ref{sec:data-concepts}.

\begin{prop}[Uniformity of differentials of a random value in the RPM] \label{hypodiff}
Let $\perm$ be a public fixed ideal random permutation of size $B$, $\clen>0$ a fixed public integer and $\vec{\Delta}=1,2,\dots,\clen$ be a fixed sequence of numbers, then consider the following procedure executed by an oracle $\co$: generate $r\leftarrow\prng$ uniformly at random, then output:
\begin{equation}\vec{Diff_r}:=\perm(r)+\perm(r+1),\perm(r)+\perm(r+2),\dots, \perm(r)+\perm(r+\clen-1) \end{equation}
Then the output of this procedure is indistinguishable from the output of a generator of random strings of length $\clen-1$ blocks, for a polynomial adversary having access to the public resource $\perm$, with respect to the security parameter $\blen$.
\end{prop}

The core idea of the proof is the following: the classical padding vector of the counter mode of encryption $[\sigma(r+1),\dots,\sigma(r+\clen-1)]$ is not safe when $\sigma^{-1}$ is made accessible to the adversary (=key exposure), since then she can invert any $\sigma(r+i)$ by querying and recover the initialization vector $r$ to distinguish a padding vector from pure randomness. So what we do in $\vec{Diff_r}$ is that we \emph{translate} this vector by a random block $v\in\F_2^\blen$. The issue is that we sample this random $v$ by querying $\sigma$ on$\dots$ $r$. But recall that $\sigma$ behaves as a random oracle (modulo avoiding collisions), thus, as long as $r$ was not queried before (nor its inverse), then $\sigma$ does output a value $v:=\sigma(r)$ uniformly at random, so this doesn't give any advantage to the adversary.

\begin{proof} We consider a cascade of games, where the view of adversary $\A$ is the same between two consecutive games up to negligible events. The first game \emph{Game$_1$} is where $\A$ faces a true random string generator, whereas the last one \emph{Game$_6$} is the actual $\vec{Diff_r}$ oracle. Let us say that, in each game, our bounded adversary $\A$ makes a total of $\quer$ queries to $\sigma$ and $\sigma^{-1}$, and of $\quer$ queries to the challenging oracle $\co$.

\emph{Game$_1$} On each query to the challenging oracle $\co$, say the $j$-th query, $\co$ sends to $\A$ a sequence of blocks $\lambda^j_1,\dots,\lambda^j_{c-1}$ sampled uniformly at random. 

\emph{Game$_2$} On each query to the challenging oracle $\co$, say the $j$-th query, $\co$ generates a sequence of random blocks $\lambda^j_0,\lambda^j_1,\dots,\lambda^j_{c-1}$ and returns to $\A$ the sequence $\lambda^j_0+\lambda^j_1,\dots,\lambda^j_0+\lambda^j_{c-1}$. The view of $\A$ is the same as in the previous game.

\emph{Game$_3$} The challenging oracle $\co$ has the same behavior as in the previous game, but the permutation oracle $(\sigma,\sigma^{-1})$ becomes "lazy": it doesn't check anymore for collisions between two outputs of different inputs, so outputs uniformly at random when queried on a new input. Such a collision event happens in the game with probability $\quer/2^\blen$, so is negligible.

\emph{Game$_4$} The challenging oracle $\co$ has the same behavior as in the previous game, but it sometimes "crashes", i.e. returns no answer. Namely, on each query, in addition to sampling the values $\lambda^j_0,\lambda^j_1,\dots,\lambda^j_{c-1}$, $\co$ also samples "for herself" a random block $r^j$ and crashes if: either one of the elements in the sequence $r^j,\dots,r^j+c-1$ (i) was already queried by the adversary to $\sigma$, or (i') overlaps with a previously sampled sequence; or, if one of the elements in the sequence $\lambda^j_0,\lambda^j_1,\dots,\lambda^j_{c-1}$ (ii) was already queried by the adversary to $\sigma^{-1}$, or (ii') overlaps with a previously sampled sequence. It is straightforward that (i) and (i') happen with probabilities $c\quer/2^\blen$, resp. roughly $2c\quer/2^\blen$ (see \cite[pf of Thm 3.30]{katz} for thinner birthday paradox estimations) because the $r^j$ are not related in any manner with the outputs given to the adversary. On the other hand, one can actually see that (ii) and (ii') also happen with probabilities $c\quer/2^\blen$ resp. roughly $2c\quer/2^\blen$, because the set of "bad" values $\lambda^j_1,\dots,\lambda^j_{c-1}$ are translated by a uniform random $\lambda^j_0$ before being given to the adversary, thus the possible sets of "bad" values are equidistributed from the point of view of the adversary. This is the core idea of the proof.

\emph{Game$_5$} The challenging oracle $\co$ has the same behavior as in the previous game except that, this time, the values $\lambda^j_{i}$ are sampled by querying the uniform $\sigma$ oracle: $\lambda^j_{i}:=\sigma(r^j+i)$. The values $\lambda^j_{i}$ are thus still sampled uniformly at random, since, outside of "crash" events, they were \emph{not queried before} to $\sigma$ (neither by $\co$ nor $\A$). Thus, the view of the adversary is the same as in the previous game.

\emph{Game$_6$} Finally, the permutation oracle $\sigma$ checks again for collisions, thus its outputs are not completely uniform anymore, introducing a negligible difference of views for the adversary compared to the previous game.
\end{proof}

\subsubsection{End of the proof of Main theorem \ref{th:mainssake}}

\emph{(2), let us consider an adversary $\A$ who is given the key $\key$}. Let us consider the SAKE game of §\ref{sub:defformalsake}: $\A$ chooses two plaintexts $\plain^{(b)}$ $b=0,1$, and is given $n-1$ shares under $\caon$ of one of them. Her view consists in $\nbs-1$ shares of the $\iv:=\iv^b$ and $\nbs-1$ fragments of $\ciph':=\ciph'{}^b$.

\emph{Game$_1$} The $\nbs-1$ shares received by $\A$ are replaced by uniformly random values.
The view is indistinguishable from the real game, because in an additive secret-sharing scheme with $\nbs$ shares, any $\nbs-1$ shares vary uniformly at random independently of the secret.

\emph{Game$_2$}  Consider this more advantageous game where the $\A$ is given \emph{all} of
\begin{equation}\ciph':=\ciph\cdot Mat^\caon\end{equation}
Since she has strictly more information in this game, her guessing advantage is bigger.

\emph{Game$_3$} Consider finally this game, where the adversary is instead given $\ciph"$, obtained from $\ciph'$ with the following successive operations (that one can also see as arising from elementary columns operations on $Mat^\caon$):
\begin{align} \cbl"_i &\longleftarrow \cbl'_i + \cbl'_{i-1} \text{ for } i=2\dots \clen-1 \text{. Overall:} \\
\ciph" :&=\ciph\cdot\begin{pmatrix}[cccccc]
 1 & 1 & 1 & 1& . & 1 \\
 1 & 0 & 0 & 0& . & 0 \\
 0 & 1 & 0 & 0& . & 0 \\
 0 & 0 & 1 & 0& . & 0 \\
 0 & 0 & 0 & 1& . & 0 \\
 . & . & . & .& . & . \\
 0 & 0 & 0 & 0& . & 1 
\end{pmatrix} \end{align}

Since the views in \emph{Game$_2$} and Game$_3$ are deduced from each other from a public invertible transformation, the guessing advantage of the adversary in \emph{Game$_3$} is larger (actually equal) than in Game$_2$.
But with the notations of Proposition \ref{hypodiff}, an adversary in \emph{Game$_3$} receives exactly
$$\vec{Diff_\iv} + \bigl(P_1^{(b)},\dots,P_\plen^{(b)} \bigr)$$
so whatever the value of the bit $b$, she has negligible advantage in distinguishing what she receives from random, by Proposition \ref{hypodiff}. So she can't a fortiori distinguish between the two possibilities.

\subsection{Revisiting the security analysis of previous works and the issue of pseudorandomness under key exposure}
\label{sec:revisiting-assumptions}  \label{sub:cexctr}

The previous scheme \cite{bib:bastion} proposes an invertible transformation which takes as input a ciphertext $\boldsymbol{y}$ output by a blockcipher $\bc$ in counter mode, which we denote $\bc$-CTR.
Given input $\boldsymbol{y}$ of size $n+1$ blocks, the transformation outputs a new chain of $n+1$ blocks, $\boldsymbol{y'}$.
The claim of \cite[Lemma 3]{bib:bastion} is that the secrecy of the plaintext of $\boldsymbol{y}$ is preserved against an adversary being leaked the key and $(n+1)-2$ out of the $n+1$ blocks of $\boldsymbol{y'}$.

The proof of security against key exposure goes through the \cite[Lemma 2]{bib:bastion}.
The proof of this Lemma 2 relies on the claim that the output of the $\bc$-CTR, $\boldsymbol{y}$, would be pseudorandom
["Recall that each block $y'_i$ is dependent on $n-1$ blocks of $\boldsymbol{y}$ and it is pseudo-random as output by the CTR encryption mode.]
However, this claim is false since the adversary knows the secret key
\footnote{As Katz-Lindell notice under their \cite[definition 3.25]{katz}, "it is meaningless to say that $\bc$ is pseudo-random if the key is known".}.
More concretely, we now show that, in the scenario considered where the key of the blockcipher $\bc$ is known, then the output of $\bc$-CTR is not random-looking \emph{even if the adversary does not know the counter}.
Note that we also sketched this problem in the idea of proof of Proposition \ref{hypodiff}.
More precisely, we consider an adversary $\A$ who chooses a plaintext $P=[P_1,\dots,P_{\clen-1}]$, is given \emph{any two} ciphertext blocks of its encryption, $\boldsymbol{C}\leftarrow \bc\text{-}\mathrm{CTR}(P)$, and is also given the encryption key, hence, knows the inverse $\bc^{-1}$ of $\bc$.
What we now show is that this adversary is able to tell apart these two blocks from a random sequence, which thus disproves the claimed pseudorandomness.
The same argument holds for any two consecutive output blocks of CBC.

More formally, denote the indices $i$ and $j$ of the blocks given to the adversary.
$\A$ is being given challenge blocks, denoted $\ti{C_i}$ and $\ti{C_j}$, and it must guess if they are a random string, in which case $\A$ should output "random"; or, if they are equal to the actual ciphertext blocks $C_i$ and $C_j$ of $\boldsymbol{C}\leftarrow \bc\text{-}\mathrm{CTR}(P)$, in which case $\A$ should output "CTR".
Before we continue, note that in the particular case where $i=0$, then the task of $\A$ is very easy (same for $j=0$).
Indeed, assuming that $\ti{C_0}$ is equal to the counter $C_0$, then $A$ should be able to recompute $\boldsymbol{C}$, of which the $j$-th block should be equal to $\ti{C_j}$.
If the case then $\A$ outputs "CTR", else, it outputs "random".

We now address the general case. Assuming that $\ti{C_i}$ and $\ti{C_j}$ are equal to the actual ciphertext blocks $C_i$ and $C_j$, then we have that there exists a certain $C_0$ such that \begin{align}
    \ti{C_i}&=P_i+\bc(C_0+i) \\
    \ti{C_j}&=P_j+\bc(C_0+j)
\end{align}
The winning strategy of $\A$ is now clear: compute $\bc^{-1}(\ti{C_i}-P_i)-\bc^{-1}(\ti{C_i}-P_i)$ and compare with $i-j$: if equal then it outputs "CTR", otherwise it outputs "random". 
This concludes the proof.

We emphasize that this absence of pseudorandomness explains why in our proof of Main Theorem \ref{th:mainssake} we needed to choose a completely different strategy and crucially relied on the RPM assumption for the $\bc_\key$ with \emph{fixed} public key $\key$.

%  The proof in \cite{bib:bastion}, follows the template of the proof of ind-CPA security for CTR mode, as done e.g. \cite[theorem 3.32]{katz}. It considers the ind-CPA CAKE game as defined in \cite[§3.2]{bib:bastion}, where $\A$ knows the key and plays against an oracle $\co$ which uses a blockcipher $\bc$ with this fixed key. Then it shows that, if such an adversary could win the ind-CPA CAKE game with non negligible advantage, then it could distinguish between the actual "pseudorandom" permutation $\bc$ and a truly random one, just as in \cite[p91]{katz}. Then it concludes a contradiction, and thus the adversary couldn't win the CAKE game. 

% Although this strategy works in \cite[theorem 3.32]{katz}%
% \footnote{"Intuitively, such a “gap” (if present) would enable us to distinguish the pseudorandom function from a truly random function. Formally, we prove this via reduction."},
% the problem here is that there is no contradiction anymore in this context of key exposure. Indeed, our adversary perfectly knows the permutation $\bc$, because she knows the key $\key$. So \emph{she can trivially distinguish it from a truly random one}. So this doesn't disprove anything about the adversary's ability to win the CAKE game. 

\subsection{Minimal sufficient hypotheses for Bastion's security} \label{sub:fixbastion}
As we just saw, public knowledge of the inverse $\bc^{-1}$ makes the output of CTR appear nonrandom. Nevertheless, we prove in Appendix \ref{sub:proofbastion} the following Proposition  \ref{prop:hypbastion}, which states that the security of Bastion holds under a certain pseudorandomness hypothesis.

Unlike the notations of \cite{bib:bastion}, we keep our indices of ciphertext blocks $C_i$ which run from $0$ to $\clen-1$, in particular we stick to our more traditional convention that $C_0$ is the initialization vector (instead of $y[n]$ in loc. cit.). We also keep our convention where $\ciph'=\ciph \cdot Mat^{Bastion}$ is the \emph{transformed} ciphertext.

\begin{prop}[Under the following hypothesis, Bastion satisfies $(\clen-2)$ CAKE security]   \label{prop:hypbastion} Sample once for all a key $\key$ which will be used in the blockcipher $\bc$ and give it to the adversary $\A$, so that she has access to $\bc$ and $\bc^{-1}$ with this key. Then, on every query of the adversary, sample a (secret) initialization vector $C_0$ uniformly at random, and for all possible tuples $(s,t)$ of distinct indices chosen by the adversary%
\footnote{The two columns that $\A$ \emph{cannot} see, which makes $\frac{\clen(\clen+1)}{2}$ possibilities depending on her choice}
in $(0,\dots,\clen-1)$, output to her the following. Choose arbitrarily an index%
\footnote{The pivot column: $u=2$ in the $7\times7$ example in the proof of Appendix \ref{sub:proofbastion}.}
$u$ distinct from $s,t$, that we give to $\A$, then output to $\A$ the vector in  $({\F_{2}}^{\blen})^{\clen-2}$ defined by:
 \begin{multline}
 \vec{Pad^{Bastion}_{(s,t,u)}} = \Bigl( \bc(C_0+s)+\bc(C_0+t)+\bc(C_0+u)\, , \\
 \bigl\{\bc(C_0+u)+\bc(C_0+i)\bigr\}_{i\in\{0,n-1\},i\notin\{s,t,u\}} \Bigr)
 \end{multline}
(where one should read, instead of $\bc(C_0+0)$, just $C_0$). Then Bastion's security is guaranteed if the adversary cannot distinguish this output from a random bit string of the same size.
\end{prop}

%% file: ROSake.tex
\section{Key exposure  resistance with \emph{no overhead} on top of encryption}  
\label{sec:analysis:keccak}

We first borrow in §\ref{sub:robaseline} a very simple ind-CPA encryption scheme based on random oracles from the paper \cite{duplex}. Then we tweak it for our purpose of secret sharing under key exposure in §\ref{sub:rodescri}, where we prove that the resulting scheme is SAKE secure. We describe an instantiation with blockciphers of two-blocks-long keys (such as AES256 for a security parameter of 128) in §\ref{sub:roaes}, and finally describe in \ref{sub:roparam} the instantiation which we implemented with Keccak, along with the choice of consistent security parameters. 

\subsection{Baseline encryption} \label{sub:robaseline}
We consider the most elementary case of the encryption scheme $\mathcal{RO}_{WRAP}$ defined in \cite[§2]{duplex}. 
\footnote{Technically we revisit the ideal vanilla case of their construction, which they call $\mathcal{RO}_{WRAP}$ in \cite[§2]{duplex}, under the simplest setting, where we consider just one single secret body $B:=\plain$. Let us recall for the interested reader that this scheme is then tweaked into authenticated Duplex mode: see their \cite[Lemma 6]{duplex}, with equivalent indistinguishability, then instantiated with an ideal sponge.}
Recall from §\ref{sec:data-concepts} that we consider a public random oracle $\co:\bigr(\F_2^{2\blen}\bigl) \lra \bigl(\F_2^\blen\bigr)^\infty $, and note $\co_{\clen-1}$ when restricting to the first $\clen-1$ output blocks.

\begin{defi}[$\mathcal{RO}_{WRAP}$] Sample once for all a secret key $\key$ uniformly at random. Then, on input $\plain$ of length $\clen-1$, sample a header $\iv$ uniformly at random and output the "one-time-padding":
$$\ciph:=\iv ||\Bigl(\co_{\clen-1}(\key||\iv)+\plain \Bigr)$$
\end{defi}

\begin{proposition} \label{prop:rowrap} $\mathcal{RO}_{WRAP}$ is an ind-CPA secure encryption scheme with secret key $\key$. \end{proposition}
This actually follows from \cite[Theorem 3.26]{katz} because the function $(\key,\iv)\lra\co_{\clen-1}(\key||\iv)$ is in particular a pseudorandom function with respect to key $\key$ and input $\iv$.

\subsection{ROSSake scheme} \label{sub:rodescri}
Our scheme ROSSake is described in the following figure. Although the fragments could be chosen completely arbitrarily, as long as they don't overlap and form a partition of the ciphertext, for simplicity we will assume as in §\ref{sec:data-concepts} that $\clen-1$ is divisible by $\nbs$, and that the fragments $F_i$ are formed by gathering $(\clen-1)/\nbs$ consecutive blocks.

\begin{figure}[htbp] \centering
		\begin{mdframed} %[ linecolor=red,linewidth=2pt, roundcorner=4pt,  backgroundcolor=olive!15, userdefinedwidth=\textwidth]
			\begin{center} \scalebox{1.0}{$\kecchare$} \end{center}
			\bigskip 
	\begin{hugenum}
        \item[Public parameters:] $0\leq t < \nbs$ some \emph{fixed} positive integers, a random oracle $\co:\F_2^{2\blen}\ra \F_2^\infty$ of arbitrary output size.
        \item[Initialization: ] Sample a fixed secret key $\key\in \F_2^\blen$ at random, which will remain the same for all inputs.
        \item[Share:] On input $\plain$ of length $\clen-1$ blocks, sample a ``header'' block $\iv\in \F_2^\blen$ uniformly at random. Then, generate the one-time-pad: $\co_\plen(\key||\iv)$. Split the padding $\plain+\co_\plen(\key||\iv)$ into $n$ disjunct fragments $F_1,\dots,F_n$. Concatenate each fragment $F_i$ with a share $\iv_i$ of $\iv$ under a $(n,n-1)$ secret sharing (e.g. additive), to obtain a complete share $S_i$.
        \item[Reconstruct:] Reconstruct $\iv$ and, knowing the key $\key$, reconstruct the one-time-pad: $\co_\plen(\key||\iv)$, then subtract it to the concatenation of the fragments $F_i$.
	\end{hugenum}
	\end{mdframed}
\setlength{\belowcaptionskip}{0pt}
\caption{RO-based key-exposure resilient secret sharing with only one round of encryption and no overhead.}
\label{pi:rocss}
\end{figure}

\begin{main} Protocol $\kecchare$ is a $\nbs-\sake$ scheme with respect to the secret key $K$. \label{th:romain} \end{main}
\begin{proof} 
(1) \emph{Let us consider an adversary who ignores the key $K$} but has all shares. Then ind-CPA of the scheme follows from Proposition \ref{prop:rowrap}. (2) \emph{Let us consider an adversary who knows the key $K$} and who plays the ind-CPA game of requirement (2) for $\nbs$-SAKE security. On every query to the challenge oracle on two plaintexts $\plain^b$, $b=0,1$, the adversary obtains $n-1$ shares of one of the chosen plaintexts $\plain^b$ generated by $\kecchare$ with the same key $\key$. [Throughout the game, the adversary can also query $\kecchare$ on any plaintext she wants.] By the property of information theoretical secret sharing, the $n-1$ shares $\iv_i^b$ given to the adversary are indistinguishable from random strings independent from all the rest of the experiment, so she can ignore them and concentrate on the $\nbs-1$ fragments $F_i^b$ that she receives. To make the argument clearer, let us give more power to the adversary and \emph{also} give to her the last missing fragment $F_i^b$ \emph{but not} the corresponding missing share $\iv_i^b$ of the $\iv^b$ (otherwise she could reconstruct the secret).

The core idea of the proof is that, in this context where the key $\key$ is exposed, the publicly know function $\co_\plen(\key||\ast):\,\iv\,\lra \,\co(\key||\iv)$
is \emph{again} a random oracle with respect to input $\iv$ (for the reader who is not convinced, we refer to the overkill domain-separation argument for random oracles done in \cite[§5]{coronhash}).

So, from the point of view of this more powerful adversary, the ind-CPA game (2) of the definition of $\nbs$-SAKE with multiple queries on the same public key $\key$, consists exactly in the following \emph{one-query} encryption game "ind1" with the following encryption challenger oracle. The oracle selects a random oracle $\co_\plen(\key||\ast)$ once for all and gives it to the adversary. Then, on every query $\plain^b$, $b=0,1$ from the adversary, the challenge oracle selects a bit $b$, samples a \emph{new} secret $\iv^b$ uniformly at random (which plays the role of a secret key in an ind1 game!) then returns to the adversary the ciphertext: $\co_\plen(\key||\iv^b) + \plain^b$.
But in particular, the function $\co_\plen(\key||\ast)$ is a pseudorandom generator in the sense of \cite[Definition 3.15]{katz}%
\footnote{As we notice in §\ref{sub:nail}, this core argument completely breaks down in general if we replace the random oracle $\co(\ast||\ast)$ by the weaker notion of a mere pseudorandom function $f(\ast,\ast)$.}
, so that we fall back in the scheme of \cite[Theorem 17]{katz}, which is proven (stronger than) ind1. Thus the adversary has negligible advantage in the game.
\end{proof}

\subsection{Can we instantiate this scheme from AES ?} \label{sub:roaes}
A convenient way to build a random oracle with input of constant size, two blocks in our use-case, consists morally in \emph{switching upside-down the construction of equation \eqref{eq:prng}} [see the comments at the end of §\ref{sub:nail}].
\footnote{This construction is suggested by the NIST: https://nvlpubs.nist.gov/nistpubs/SpecialPublications/NIST.SP.800-90Ar1.pdf}
Let us formalize this in the ideal blockcipher model.
\begin{proposition} \label{prop:prng} Consider an ideal blockcipher $\bc^2$. Let us consider a fixed public counter, e.g. $1,2,3,\dots$. Then the following function is a random oracle with input size the keysize $|\mathcal{L}|$ of $\bc^2$:
\begin{equation} \text{on input $\mathcal{L}$, return: } \bigl[\bc^2_{\mathcal{L}}(1), \bc^2_{\mathcal{L}}(2), \bc^2_{\mathcal{L}}(3),\dots\bigr] \end{equation}
\end{proposition}
Importantly, for our purpose we thus use an ideal blockcipher $\bc^2$ with key $\mathcal{L}$ of size \emph{two blocks}: twice larger than the key $\key$ of the final $\kecchare$ scheme obtained. Thus the exponent $2$, to differentiate it from the $\bc$ of the previous $\caon$ scheme of keysize one block.

\begin{proof} The proposition results from the definition of an ideal blockcipher. Indeed for each new input $\mathcal{L}$, then a fresh new random permutation oracle $\bc^2_{\mathcal{L}}$ is created, so that the outputs on $\bc^2_{\mathcal{L}}$ when queried on the counter $1,2,\dots$ are uniform and \emph{independent} from the outputs of previously queried oracles $\bc^2_{\mathcal{L}'}$ with other keys ${\mathcal{L}'}$.
\end{proof}
We hid under the carpet that the distribution of outputs is not totally uniform, since the permutation oracle $\bc^2_{\mathcal{L}}$ avoids collisions (e.g. will select the output of $10$ outside of the previous outputs of $1,2,\dots,9$). But this bias, of order of magnitude $\clen/2^\blen$ is by definition negligible. For example if the oracle of Proposition \ref{prop:prng} were used in place of the construction \eqref{eq:prng} to produce random-looking strings, then this bias would gives no more than a negligible advantage to an adversary that makes a number of queries negligible compared to $2^\blen$. A formal argument for this detail can be found along the lines of the proof of the PRF/PRP Lemma, see e.g. \cite[§7]{maurerindiff}.

In conclusion, for our purpose in the case where the security parameter is one block of size 128 bits, then AES256 has indeed key of size two blocks (256), thus by the previous proposition it does the job in the ideal blockcipher model .

\subsection{Our instantiation with Keccak} \label{sub:roparam}
Assuming the existence of a fixed random permutation $\sigma$, the authors of Keccak could prove the existence of a random oracle of arbitrary input and output size in \cite{keccak} (adopted for SHA3). From it they could build an encryption scheme "SpongeWrap" which satisfies a pseudorandomness property. Namely, \cite[Theorem 1]{duplex} can be read in the light of \cite[Definition 3.28]{katz} as stating that, under the RPM, then SpongeWrap is indistinguishable from very large keyed pseudorandom permutation (from its input domain onto its image).

For mere privacy (not authentication) concerns, the security parameters of \cite{duplex} are $\keys$ the length of the key and $c$ the "capacity" of the sponge ($c$ is not the ciphertext's size in \cite{duplex}!). The first formula of \cite[Theorem 1]{duplex} then states that for an adversary allowed to do a negligible number of queries (noted here $N$ and $q$) with respect to $2^c$ and $2^\keys$, then the distinguishing advantage is essentially in $\max(2^{-c},2^{-\keys})$. For the sake of comparison with the schemes Bastion and $\caon$ that we instantiated with AES128, we thus took security parameters $|\iv|=\keys=c=128$ (and bit rate $r=1600-c$) for Keccak in our simulations in \ref{sec:comparison:speed}. We set the number of rounds of the permutation $f$ to $24$, as recommended by the guidelines%
\footnote{Note on Keccak parameters and usage, NIST hash forum, 2010}.

In conclusion: Keccak might seem overkill to instantiate a random oracle with \emph{constant} small input size, but it is our preferred choice since it gave as good performances than blockciphers, plus the Duplex mode also enables an authentication of the ciphertext with no additional round of encryption, which is interesting for integrity/robustness issues of secret sharing.

%% file: comparison.tex
\section{Comparison with Relevant Works}
\label{sec:comparison}

In Table ~\ref{tab:comparison} we compare  SSAKE and $\kecchare$ schemes with relevant works in terms of complexity, memory size, with regard to their compliance with our security properties. After commenting Table~\ref{tab:comparison} and detailing how the selected schemes were implemented, we present their experimental throughput in Figure~\ref{fig:performance}.  

\renewcommand{\arraystretch}{1.3}
\begin{table*}[h]
  \caption{\textit{Comparison with relevant works in terms of number of required encryption rounds, number of overhead XORs required during the post-processing, total size of the stored data, \textbf{(1)} ind-security given all the shares, CAKE security, and SAKE security.}}
  \label{tab:comparison}
  \begin{threeparttable}[t]
  \begin{tabular}{l|l|l|l|l|l|l}
    \toprule
    Algorithm & Enc. rounds & Extra XORs & Total data size & \textbf{(1)} ind-CPA & CAKE & SAKE \\
    \midrule
    CTR Enc. & 1 & N/A & $\ciphl$ & Yes & 1 CAKE  & No \\
    
    SSMS & 1  & N/A & $\ciphl+\nbs\keys$ & No & No & No \\
    
    Rivest AONT & 2 & 1 & $\ciphl+\keys$  & No & No &  No \\
    
    Desai AONT & 1  & $\clen$ & $\ciphl+\keys$ & No & No &  No  \\
    
    Rivest AON & 3  & 1 & $\ciphl+\keys$ & Yes & $\clen-1$ CAKE &   $\nbs$-SAKE  \\
    
    Desai AON  & 2  & 2 & $\ciphl+\keys$ & Yes & $\clen-1$ CAKE & $\nbs$-SAKE  \\
    
    Bastion & 1 & $2\clen$ & $\ciphl$ & Yes & $\clen-2$ CAKE &  $\nbs$-SAKE  \\
    
    \textbf{\caon} & 1 & $\clen + \nbs - 1$ & $\ciphl+(\nbs-1)\keys$ & Yes & $\nbs-1$ CAKE &  $\nbs$-SAKE \\
    
     \textbf{\kecchare} & 1 & $\nbs$ & $\ciphl+(\nbs-1)\keys$ & Yes & $\nbs-1$ CAKE & $\nbs$-SAKE \\
    
  \bottomrule
\end{tabular}
 \begin{tablenotes}
\item[1] Note that the keysize $|K|=|B|$ becomes negligible as $|\ciph|$ increases in size.
%       \item[1] Unlike in CAKE secure schemes, in \caon the way of dispersal of the blocks over share is important and thus its CAKE security varies depending on these dispersal.
%       \item[2] Scheme resistant to key exposure under the assumption that the encryption key stored within the data was correctly generated.
%      \item[3] The $(\nbs-1)$ SAKE property of $\caon$ scheme holds under a lighter assumption than the other schemes satisfying both Security (1) and (2).
   \end{tablenotes}
\end{threeparttable}
\end{table*}

\subsection{Complexity and storage} 

We compare the $\caon$ and $\kecchare$ schemes in terms of complexity with relevant works: Krawczyk's SSMS, Rivest's and Desai's AONT and AON, Bastion. The CTR encryption is used as a baseline. 
The important difference between the new generation of schemes protecting against key exposure and former algorithms is the number of encryption rounds. SSAKE, $\kecchare$, and Bastion use only one single encryption round while Rivest's and Desai's AON apply two or more rounds.

Note that, Krawczyk's SSMS and Desai's AONT use a single round but do not protect against key exposure: see §\ref{sub:krawbreak} for a discussion. Whereas Rivest's and Desai's AON are CAKE, but require two encryption rounds: see §\ref{remark-on-r}. Finally see §\ref{sub:sake-cake} for how a $\clen-2$ CAKE scheme is $\nbs$-SAKE for free, provided $\nbs\leq 2\clen$, which explains our last $\sake$ column. See also §\ref{sub:sake-cake} for the direction $\nbs$-SAKE to $\nbs-1$ CAKE, which explains the $\nbs-1$ CAKE entries for our schemes.

Both Bastion and SSAKE use a linear transform post-processing on top of encryption. The difference is that SSAKE transform uses twice less XOR than the Bastion's one. Bastion scheme applies a linear transform over the encrypted data requiring $2c$ XORs. $\caon$ applies only a linear transform (one pass of XORs) over the ciphertext and an additive PSS over the $\iv$ (that has a negligible impact: $n$ xors, on the scheme performance when $\clen$ is large). 

$\kecchare$ does not require to apply a transform over the whole ciphertext. It needs only $\nbs$ XORs more in addition to data encryption, which is negligible as $\nbs$ is no more than 20-30 (when shares are distributed over multiple servers inside a data center, in a multi-cloud scenario it would be no more than 4).

The schemes of Rivest's, Desai's and \cite{bib:bastion} produce a total stored data (= all the shares) of a size equal to the ciphertext's $\ciphl$. Whereas the schemes SSMS, $\caon$, and $\kecchare$ require an additional storage space of $\nbs$ or $\nbs-1$ additional blocks (=secret sharing of the key in SSMS or of the $\iv$ in our scheme), which is a constant with regard to the size of the original data. Therefore, this overhead is negligible for large files.

 %As a baseline, we use CTR encryption that requires $c-1$ block cipher operations and $c-1$ exclusive-or operations. Rivest's and Desai's AON doubles (Desai) or triples (Rivest, since the hash of ciphertext is computed) the number of block ciphers operations in comparison to normal data encryption. Bastion scheme applies only a linear transform over the encrypted data requiring $2c$ XORs. $\caon$ applies only a linear transform (one pass of XORs) over the ciphertext and an additive PSS over the $\iv$ (that has a negligible impact: $n$ xors, on the scheme performance when $\clen$ is large). Both $\caon$ and Bastion require an additional pass of xors, used to add the counter in CTR mode of encryption. 

\subsection{Performance evaluation}
\label{sec:comparison:speed}

%In Figure~\ref{fig:performance} we present results of an experimental evaluation. First, some details about how they are implemented and assessed are provided in §\ref{sub:impl}. Then in §\ref{sub:experiment}, few comments are given relatively to figure~\ref{fig:performance}. 

 Relevant algorithms were implemented using the same programming style in JAVA with JDK 1.8 on DELL Latitude E6540, X64-based PC running on Intel\textsuperscript{\textregistered} Core\textsuperscript{TM} i7-4800MQ CPU @ 2.70 GHz with 8 GB RAM, under Windows 7. Standard $javax.crypto$ library was used and the official Keccak implementation was used for Lake Keyak~\footnote{https://keccak.team/keyak.html} to instantiate $\kecchare$: we refer to §\ref{sub:roparam} for our choices of consistent security parameters. A random data sample was used for each measurement and each presented result is an average of 30 measurements. AES-CTR-128 was used as the algorithm for symmetric encryption. AES-NI was enabled. Results are somewhat consistent with those presented in~\cite{bib:bastion} when taking into account the difference between AES and AES-NI (a factor of 3 in performance was observed in our implementations) as well as differences between hardware platforms. 

\begin{figure*}
\includegraphics[width=0.99\linewidth]{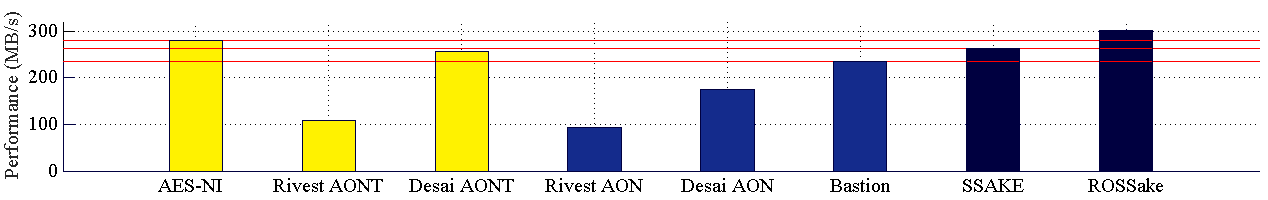}
\caption{\textit{Performance comparison in MB/sec (in yellow: schemes with only (1) security property, in blue: schemes with both (1) and (2) security properties, dark blue: our schemes). Rivest's, Desai's, Bastion, and SSAKE scheme use AES-NI as the underlying symmetric key encryption. The performance of the SSMS scheme is essentially equal to the one of the pure data encryption (AES-NI). SSAKE reduces by half the performance overhead in comparison to Bastion. ROSSake based on Lake Keyak comes with no overhead and is the fastest scheme.}}
\label{fig:performance}
\end{figure*}

As shown in the Figure \ref{fig:performance}$, \caon$ is the fastest among schemes protecting encrypted data against key exposure based on mainstream symmetric encryption with AES. Protection against key leakage is achieved with an overhead of only 7\% against a simple data encryption. The second fastest scheme with this respect, Bastion, results in an overhead of around 19\% in comparison to data encryption. Our scheme $\kecchare$, implemented with the less conventional Keccak-based encryption, turns out to be the fastest of all $\sake$ schemes. 

%% file: conclusions.tex
\section{Conclusions}
\label{sec:conc}

In this paper we revisited computational secret sharing aiming at (1) protecting data against an adversary possessing all the shares but not the encryption key, (2) protecting data against an adversary possessing the encryption key and all but one share, and at the same time (3) respecting performance, storage occupation, and scalability requirements. We introduced a new security model inspired by the CAKE model and adapted to take into account the dispersion of shares in a distributed environment. The privacy threat under key exposure is defined in terms of the number of revealed shares and not, as so far in the all-or-nothing literature, the number of revealed ciphertext blocks. We presented a versatile computational secret sharing scheme - $\caon$ - that verifies both security properties \textbf{(1)} and \textbf{(2)} of this new model. Complexity and empirical evaluations show that it performs faster than the fastest known relevant scheme - Bastion - since the linear overhead is reduced by half, achieving the point (3) above. We provided a detailed security analysis of not only the presented scheme but also of previous relevant work. We finally presented and implemented an alternative random-oracle based scheme ---that can be instantiated with blockciphers (such as AES256 for a security parameter of 128) or e.g. with Keccak--- which requires no processing overhead on top of encryption and turned out to be the fastest of all.

\subsection*{Acknowledgments}
{\small We would like to thank prof. Srdjan Capkun for inviting us to a seminar at ETH Zurich and for a helpful discussion on the topic. We thank Marc Stevens, Mohamed Tahar Hammi and Han Qiu for very fruitful discussions. We would also like to thank the journalist Ingrid Fadelli for inviting us to do an interview for TechXplore%
\footnote{\url{https://techxplore.com/news/2019-02-circular-all-or-nothing-approach-key-exposure.html}}. The second author was (partially) funded by ANR-15-CE39-0013-01 “Manta”.
\noindent Some of the results in this paper were presented in December 2018 at a seminar at LINCS Paris.
}
%In the future we would like to provide a fine-grained implementation of SSAKE that would make the performance overhead negligible in comparison to simple data encryption.

%% file: appendix.tex
\begin{appendix}
 
 \section{Proof of Proposition \ref{prop:hypbastion}}
 \label{sub:proofbastion}
Let us recall the matrix of the linear transform of Bastion, here for a transformed $\ciph'$ of size $\clen=7$:
$$\ciph'=\ciph\cdot \begin{pmatrix}[cc|ccccc] 0 & 1 & 1 & 1 & 1 & 1 & 1\\
1 & 0 & 1 & 1 & 1 & 1 & 1\\
1 & 1 & 0 & 1 & 1 & 1 & 1\\
1 & 1 & 1 & 0 & 1 & 1 & 1\\
1 & 1 & 1 & 1 & 0 & 1 & 1\\
1 & 1 & 1 & 1 & 1 & 0 & 1\\
1 & 1 & 1 & 1 & 1 & 1 & 0
\end{pmatrix}$$

The last $\clen-2$ columns are separated from the two first to illustrate the view of an adversary who would always asks to see $\ciph'_{\A}=(C'_2,\dots,C'_{\clen-1})$ during the CAKE game. Performing elementary columns operations inside the $\clen-2$ columns of the adversary, like in the proof of $\caon$ (§\ref{sub:proofcaon}):
\begin{align} C"_{i\geq 3}:=C'_i+C'_2 \\
C"_2:=C'_2+\sum_{i\geq 3} C"_i
\end{align}
we obtain the adversary's view on the right of the vertical separator:
\begingroup
\renewcommand*{\arraystretch}{.5}
$$\ciph"=\ciph\cdot\begin{pmatrix}[cc|ccccc]
0 & 1 & 1 & 0 & 0 & 0 & 0\\
1 & 0 & 1 & 0 & 0 & 0 & 0\\
1 & 1 & 1 & 1 & 1 & 1 & 1\\
1 & 1 & 0 & 1 & 0 & 0 & 0\\
1 & 1 & 0 & 0 & 1 & 0 & 0\\
1 & 1 & 0 & 0 & 0 & 1 & 0\\
1 & 1 & 0 & 0 & 0 & 0 & 1
\end{pmatrix}$$
\endgroup

\vfill \eject 

which is the equivalent to the former, from the point of view of an adversary seeing only the last $(\clen-2)$ columns: $\ciph"_{\A}=(C"_2,\dots,C"_{\clen-1})$. (Let us repeat the argument of §\ref{sub:proofcaon}: from this new view one can deduce the former one, so the guessing advantage of the adversary is larger ---actually: equal--- than with the previous view). As in §\ref{sub:proofcaon}, $\ciph"_{\A}$ is equal to a value known to the adversary (deduced from $[0,\plain^{(0)}]$ or $[0,\plain^{(1)}]$ by transforming them according to the previous matrix), plus $\overrightarrow{Pad^{Bastion}_{(1,2,3)}}$, which is assumed indistinguishable from a random uniform vector. Finally, we must loop over all the possible $(\clen-2)$ uples of columns among $n$ that $\A$ chooses to see. So we end up with asking for the $\clen(\clen+1)/2$ distinct indistinguishability assumptions of vectors $\overrightarrow{Pad^{Bastion}_{(s,t,u)}}$, as formulated in Proposition \ref{prop:hypbastion}.

\section{More on AONT's}
\label{sub:aonts}

Rivest's \cite{rivest} introduced the first known AONT, which is meant to be a pre-processing step applied before data encryption. During this AONT, input data is encrypted into ciphertext $\ciph$  using a first random key $K_{1}$. A hash of the encrypted data is then computed, XOR-ed with $K_{1}$, and appended as the last block of the ciphertext $C_{c}=Hash(\ciph)+K_{1}$. After such data transform, it is not possible to obtain the right hash, and consequently the encryption key, without possessing the whole ciphertext. However, an attacker knowing $K_{1}$ can obviously decrypt a fragment of the ciphertext in her possession. Therefore, Rivest suggests to re-encrypt the transformed data one more time with a different encryption key $K_2$. The data will be then protect against the exposure of the first  $K_1$ or second key $K_2$ (but not both of them) unless the whole ciphertext is revealed. 

Later, Boyko formalized the all-or-nothing transform \cite[Definition 2]{boyko}. The numerous subsequent variants of AONT are encompassed by the following loose definition: it consists of a randomized map $AONT$, which maps an input $\x$ of \emph{fixed size} $\xlen$ bits to an output $\y$ of \emph{fixed size} $\ylen$ bits and such that a polynomial adversary $\A$ has negligible advantage in the following indistinguishability game, where ${L}$ is a set of missing bit positions in $\y$. "Negligible" is to be understood with respect to the security parameters: length of a certain symmetric key $k_s$ (underlying to $AONT$) and number of missing bits $|{L}|$.

\begin{itemize}
    \item  $\A$ has access to $AONT$ and is given ${L}$ (or possibly adaptively chooses ${L}$). She outputs two plaintexts $\x_0$ and $\x_1$.
    \item $\A$ is given ${AONT(\x_b)}$, with bit positions $L$ missing, and $b\in\{0,1\}$ is a random index unknown to the adversary.
    \item To win the game, $\A$ has to successfully guess the value of $b$.  
\end{itemize}

\end{appendix}